\documentclass[sigconf]{acmart}
\usepackage{subfigure, multirow}
\usepackage{balance}
\def \etal {et al. }
\AtBeginDocument{%
  \providecommand\BibTeX{{%
    \normalfont B\kern-0.5em{\scshape i\kern-0.25em b}\kern-0.8em\TeX}}}

\setcopyright{acmcopyright}
\copyrightyear{2020} 
\acmYear{2020} 
\setcopyright{acmcopyright}\acmConference[ICMI '20 Companion]{Companion Publication of the 2020 International Conference on Multimodal Interaction}{October 25--29, 2020}{Virtual event, Netherlands}
\acmBooktitle{Companion Publication of the 2020 International Conference on Multimodal Interaction (ICMI '20 Companion), October 25--29, 2020, Virtual event, Netherlands}
\acmPrice{15.00}
\acmDOI{10.1145/3395035.3425202}
\acmISBN{978-1-4503-8002-7/20/10}

\begin{document}
\fancyhead{}
\title{Training Strategies to Handle Missing Modalities for Audio-Visual Expression Recognition}


\author{Srinivas Parthasarathy}
\affiliation{%
	\institution{Amazon}
	\streetaddress{905 11th Ave, CA 94089}
	\city{Sunnyvale}
	\state{California}
	\postcode{94089}}
\email{parsrini@amazon.com}

\author{Shiva Sundaram}
\affiliation{%
	\institution{Amazon}
	\streetaddress{905 11th Ave, CA 94089}
	\city{Sunnyvale}
	\state{California}
	\postcode{94089}}
\email{sssundar@lab126.com}


\begin{abstract}
	
	Automatic audio-visual expression recognition can play an important role in communication services such as tele-health, VOIP calls and human-machine interaction. Accuracy of audio-visual expression recognition could benefit from the interplay between the two modalities. However, most audio-visual expression recognition systems, trained in ideal conditions, fail to generalize in real world scenarios where either the audio or visual modality could be missing due to a number of reasons such as limited bandwidth, interactors' orientation, caller initiated muting. This paper studies the performance of a state-of-the art transformer when one of the modalities is missing. We conduct ablation studies to evaluate the model in the absence of either modality. Further, we propose a strategy to randomly ablate visual inputs during training at the clip or frame level to mimic real world scenarios. Results conducted on in-the-wild data, indicate significant generalization in proposed models trained on missing cues, with gains up to 17\% for frame level ablations, showing that these training strategies cope better with the loss of input modalities.

\end{abstract}


\maketitle
\vspace{-0.1cm}
\section{Introduction}
Expressions are a fundamental part of human communication. Humans inherently process and react to situations depending on how well they detect others expressions. Automatic recognition of expression therefore plays an important role in building intuitive interfaces for human-machine interaction. Expression recognition systems have varied uses in fields such as health \cite{Cummins_2015}, security \cite{Clavel_2008}, customer interaction \cite{gupta2007two}. In particular, expression recognition has interesting uses in communication services where systems can detect the onset of boredom \cite{jang2015analysis}, extract salient regions (hotspots) in an interaction \cite{Parthasarathy_201x} or indexing \cite{richard2013overview}

Recently, the proliferation of audio and visual sensors has seen a rise in development of expression recognition systems that are trained on data from the wild. Cai \etal provide strategies to fuse audio and visual features at the feature and model level \cite{cai2019feature}. They use an ensemble architecture to extract visual cues from the face and combine with audio. For feature level fusion they use a simple concatenation of features and for model level fusion they employ a Bayesian network, showing competitive results for the task. Kollias \etal \cite{kollias2019expression} provide a late fusion approach to concatenate audio and visual cues extracted through a deep convolutional neural network. Vielzuf \etal study the temporal fusion of audio and visual cues using 2D and 3D CNNs. \cite{vielzeuf2017temporal}. Avots \etal study cross-corpus evaluation of audio-visual expression recognition systems, combining three in the wild datasets for their training \cite{avots2019audiovisual}. Other works have proposed ensemble approaches for expression recognition in the wild \cite{hu2017learning, gideon2016wild}. These studies have established that the audio and visual cues provide complimentary information for expression detection. Furthermore, Albanie \etal successfully transferred emotion information from a teacher model trained on visual cues to a student using audio cues \cite{albanie2018emotion}. While these systems, attempt to bridge the gap between laboratory and realistic settings, another particular challenge arises when either input modality is completely or partially missing. Complete ablation of a modality could stem from user-initiated loss of modality, e.g. turning off the camera and partial ablations could stem from connectivity issues where frames or sequence of frames are dropped from a session. A significant question that needs further research is model robustness in the absence of input modality.


Building upon these works, this study analyzes the performance of audio-visual expression recognition systems when the input modalities are missing. While previous works have studied the effect of input modalities for other audio-visual tasks, for e.g. lip-reading \cite{Chung_2017}, action recognition \cite{Lee_2019}, this study understands their dynamics for expression recognition. We conduct evaluations on the Aff-Wild2 database \cite{kollias2018aff}. We first establish a baseline state-of-the-art transformer architecture with a cross-modal attention layer  \cite{Vaswani_2017attention}. This self-attention model intuitively learns to focus on important parts of the input signal. We illustrate the degradation in performance of this system when modalities are ablated, showing its susceptibility to the loss of inputs. To combat this, we also propose simple techniques to ablate data during training. We a) Ablate data at the clip level where the data is zeroed for the entire clip, b) Ablate data at the frame level where frames are either replaced by zeros or repeated. The training and evaluation strategies represent real communications scenarios that span entire loss of a modality or intermittent losses. We provide comprehensive analysis of all models evaluating them at various ablation operating points. Our results show that proposed models significantly outperform the baseline model under ablated test conditions both at the frame and clip level, while achieving near to state-of-the-art performance when tested under ideal conditions, indicating a superior generalization of the proposed models. 

This paper mainly contributes by evaluating audio-visual expression recognition systems under a wider range of mismatched conditions which are present in real world scenarios.





\begin{figure}[t]
	{
		\includegraphics[width=5.4cm]{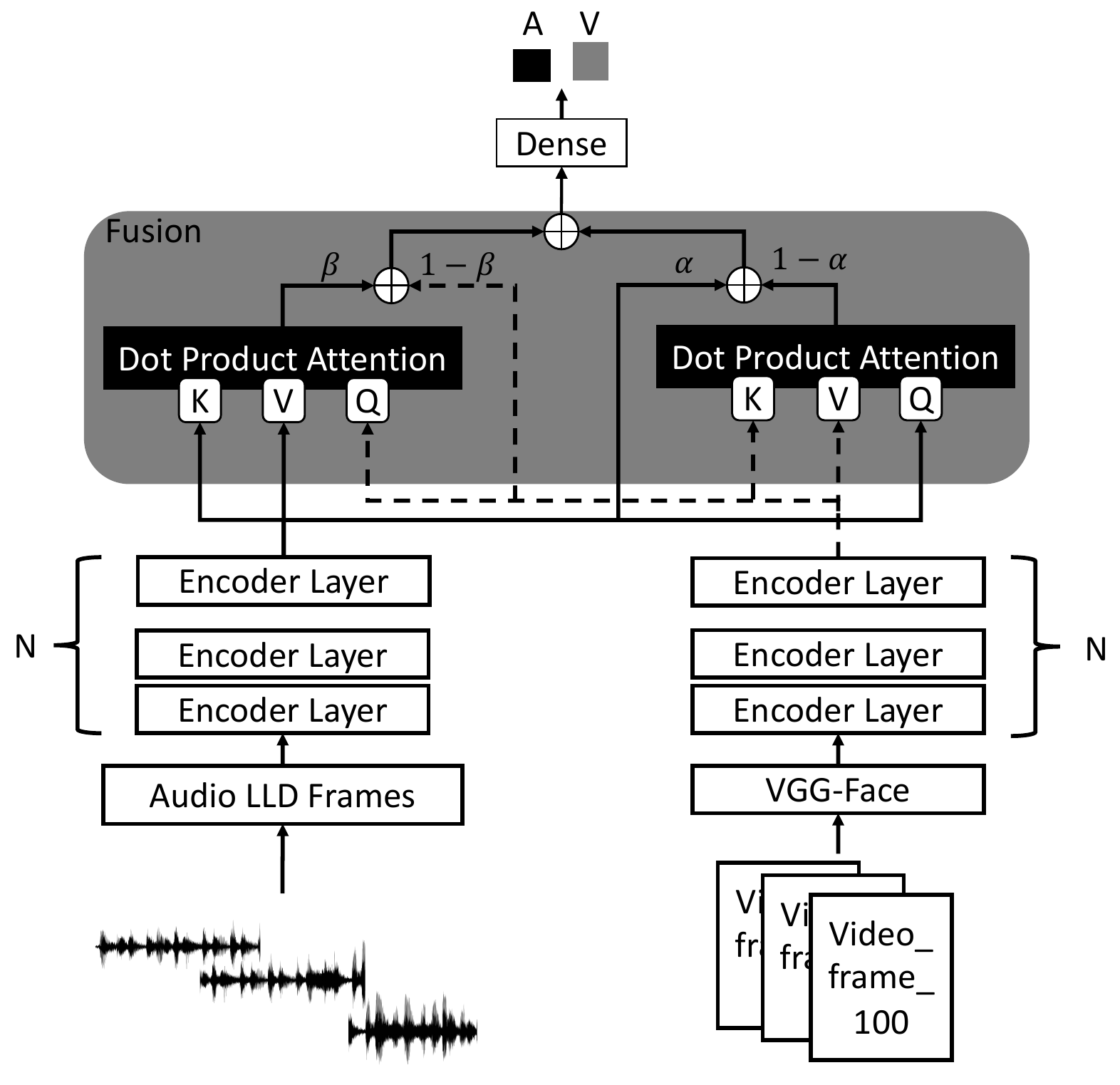}
	}
	\vspace{-0.1cm}
	\caption{Baseline transformer architecture with a cross-modal attention layer for audio-visual expression detection}
	\label{fig:baseline}
	\vspace{-0.5cm}
\end{figure}

\begin{figure*}[t]
	\subfigure[Clip zeroing]
	{
		\includegraphics[width=5cm]{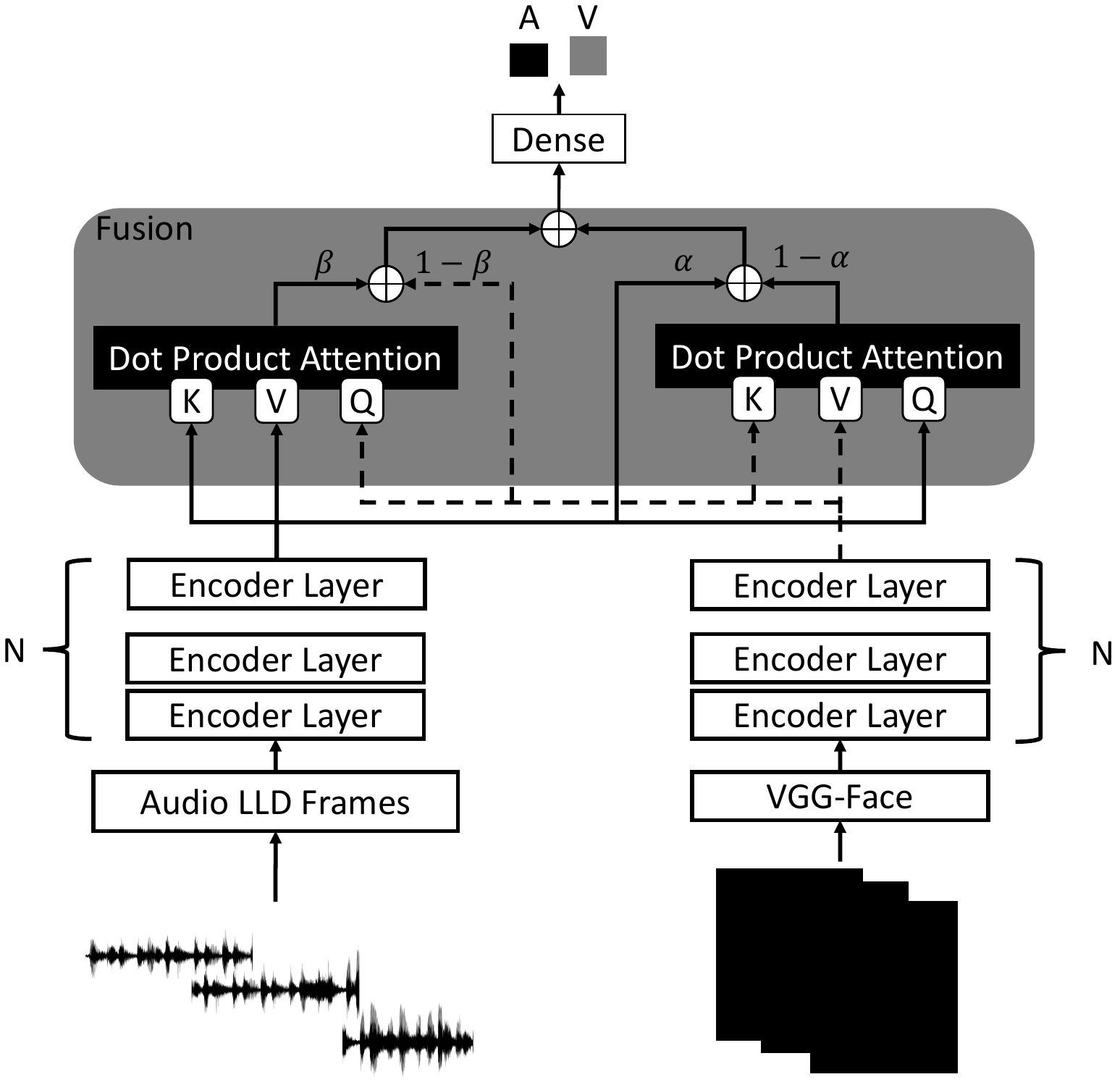}
		\label{fig:clip}
	}
	\subfigure[Frame zeroing]
	{
		\includegraphics[width=5cm]{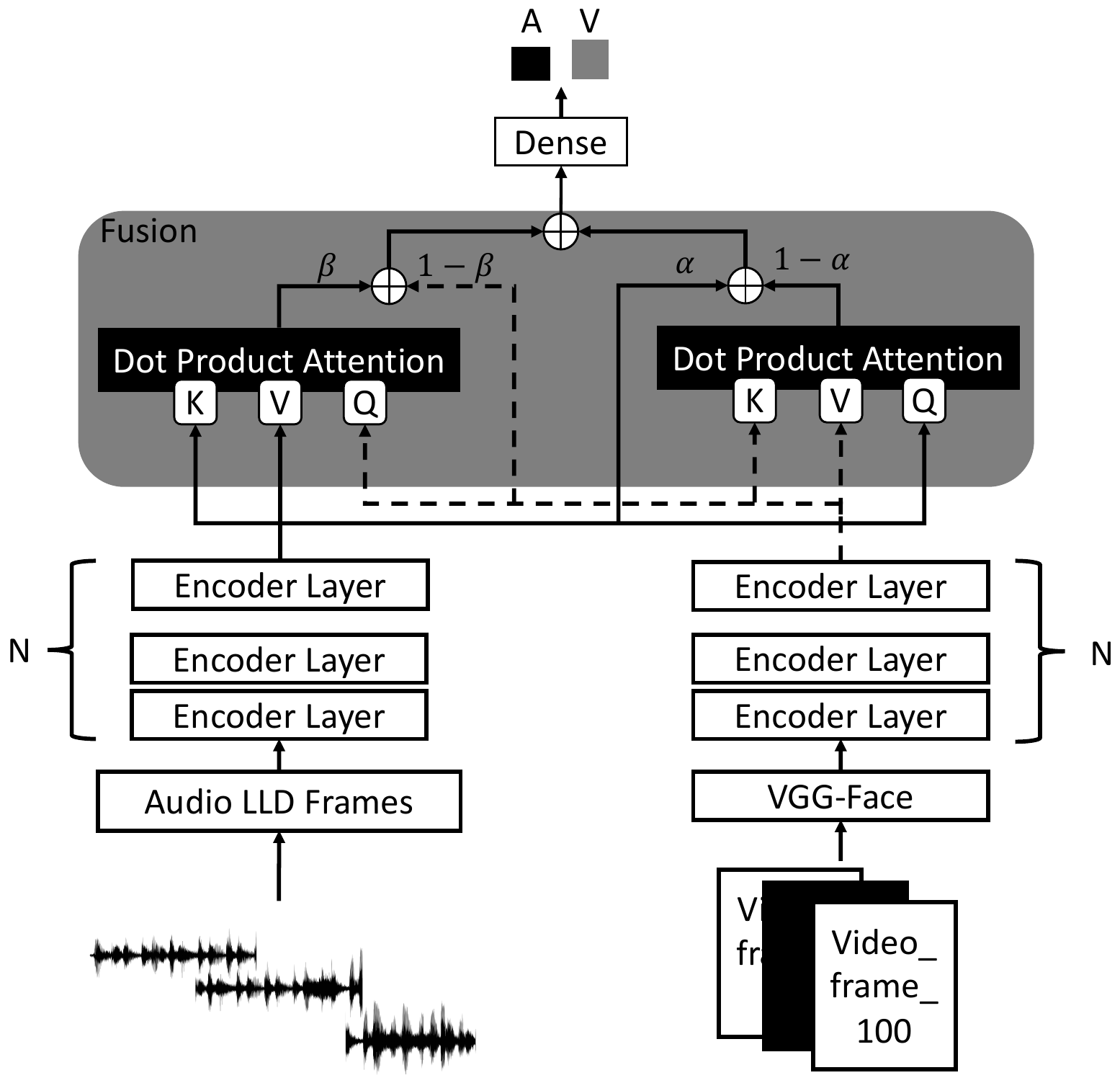}
		\label{fig:frame_zero}
	}
	\subfigure[Frame level repetition]
	{
		\includegraphics[width=5cm]{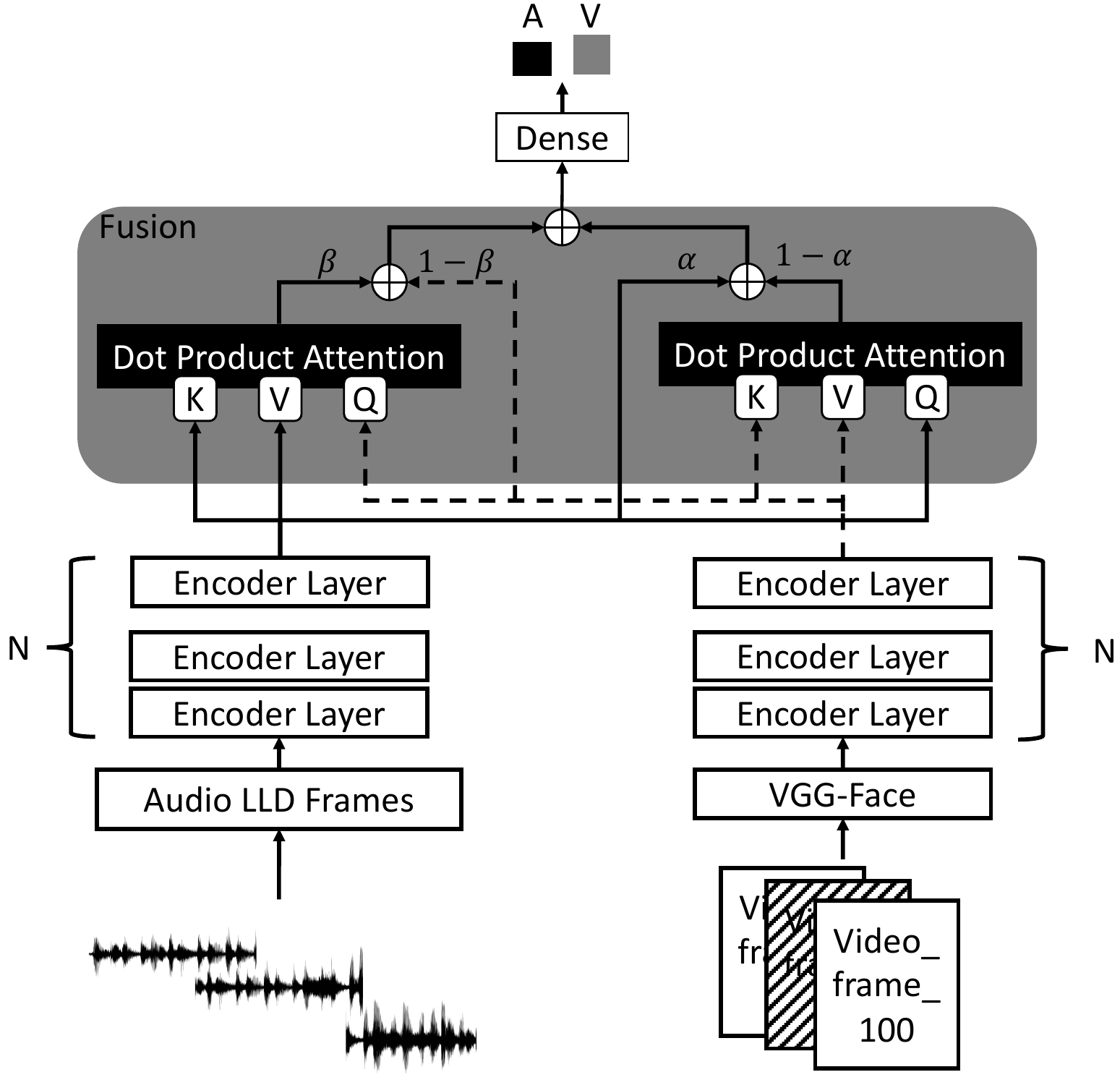}
		\label{fig:frame_repetition}
	}
	
	\caption{Proposed ablation strategies. All ablations are conducted on the visual modality}
	\vspace{-0.3cm}
	\label{fig:rep}
\end{figure*}

\vspace{-0.2cm}

\section{Proposed Model}
This work focuses on the time-continuous recognition of expressions. Given a sequence of audio-visual frames, the task is to predict the arousal (level of excitement) and valence (level of positivism) values for each frame in the sequence. 
\vspace{-0.2cm}
\subsection{Crossmodal Transformer Architecture}
This study uses a transformer architecture \cite{Vaswani_2017attention} as a baseline for audio-visual expression recognition (Figure \ref{fig:baseline}). Few previous studies have explored the use of transformers for expression recognition \cite{li2019improved, rahman-etal-2020-integrating}. First, the audio and video inputs are independently passed through several encoder layers in two separate branches, one for the audio and one for the video. The encoder layers consist of dot-product attention layers, attending to significant regions within each modality. The dot-product attention layers work by constructing \emph{Key} (K), \emph{Value} (V) and \emph{Query}(Q) triplet. The output is then calculated as 
\begin{equation}
Y = \sigma(KQ^T)V
\label{eq:att}
\end{equation}
where $\sigma$ denotes the softmax operation.

We  fuse the audio and video cues using a cross-modal attention layer. This layer contains two dot-product attention components that attend to the other modality. Similar to the encoder layers, the output is obtained using Equation \ref{eq:att}, with the main difference being K and V are obtained from one modality while Q is obtained from the other modality. The cross-modal attention outputs are added back to the corresponding encoder representations using learnable scalar weights $\alpha$ and $\beta$. The overall representation from the cross-modal layer is the sum of the individual components. A dense layer is then used to get output predictions. This cross-modal attention architecture has shown competitive performance for automatic speech recognition tasks \cite{paraskevopoulos-etal-2020-multimodal}.

\subsection{Training with missing data}

To better cope with the loss of input modality, this study proposes two strategies to train with missing data (Figure \ref{fig:rep}). Our results with the baseline model (Section \ref{sec:results}) indicate that the performance is more severely affected by the loss of visual cues than  and the absence of audio cues negligibly affects the performance. Therefore the focus of the paper is to train by ablating the visual modality. In the first method we ablate data at the clip level - \textbf{Clip-Zero.} Each clip is zeroed out with a given probability $p_{clip}$. This corresponds to zeroing out an entire sequence of frames. In the second method we ablate data at the frame level. At the frame level, we further subdivide the training into two categories. Each frame in a sequence is either zeroed out \textbf{Frame-Zero} or repeated \textbf{Frame-Repeat} with a probability $p_{frame}$. Note, that these training techniques can be viewed as special forms of data augmentation. Frame-zero and Clip-zero can be viewed as extreme cases of input feature dropout where the entire input at the frame or sequence level is dropped out. With these viewpoints, we expect the model to better generalize to different test conditions. Overall these training strategies imitate the different scenarios in which data might be missing during evaluation and real use cases.

\vspace{-0.2cm}
\section{Experimental Setup}

\subsection{Dataset}
This study uses the the Aff-Wild2 database \cite{kollias2018aff}. The dataset consists 558 videos sourced from YouTube, mostly containing single person expressions. Videos are captured in the wild conditions with hugely varying acoustic and visual environments, making it ideal for this study. All the videos are captured at a frame rate of 30 \emph{frames per second} (fps) with around 2.8 million video frames in the entire dataset. The dataset is annotated for 3 main behavior tasks (basic expression classification, valence-arousal estimation and action unit detection.) This paper focuses on frame level valence-arousal estimation. Each frame in the dataset is annotated  by 4 or 8 annotators for valence and arousal score in a continuous scale ranging between [-1, 1]. Further, the dataset is divided into speaker independent train, validation and test subsets consisting of 350, 70 and 138 videos respectively. 

\vspace{-0.2cm}
\subsection{Features and Preprocessing}
\subsubsection {Video}
 Following previous studies, we extract bounding boxes for to detect frontal faces every video frame  \cite{kollias2019expression}. The face detector is based on an SSD face detector with a ResNet architecture  \cite{Liu_2016}. The face bounding boxes are further cropped and resized to dimensions of 96x96x3. Pixel intensities are normalized between [-1, 1]. To extract features from the face frames, we use a pretrained VGG-Face model \cite{Parkhi15}. Representations from the first fully connected layer with a dimension of 4096 is fed as input to the video branch of the transformer.

\vspace{-0.2cm}
\subsubsection {Audio}
For the audio features we use the features introduced at the para-linguistic challenge in Interspeech 2013 \cite{Schuller_2013}. These features have shown superior performance for audio expression recognition tasks. This feature set consists of a variety of acoustic features grouped into spectral, energy related and voice features such as \emph{Mel Frequency Cepstral Coeffecients} (MFCC), loudness and fundamental frequency (F0). Features are extracted over 30ms windows with a 10ms step. The feature set consists of 65 \emph{low-level descriptors} (LLD) and Z-normalized over the train set. To synchronize the audio and video features the audio frames are further downsampled by a factor of 3.3 and concatenated over a 2s window to increase context and match the dimension of the video branch. Overall, the audio branch of the transformer receives a 3900-dimensional vector.

\vspace{-0.25cm}
\subsection{Experimental framework}
Arousal and valence predictions are made on sequences of 100 frames which corresponds to 3s of the input video. Audio and video encoders branches of the transformer contain 2 self-attention layers with 512 nodes and 4 heads for multihead attention. For the proposed models, the probability of ablation at both clip level $p_{clip}$ and frame-level $p_{frame}$ is fixed at 0.5. Since the test data set is not released publicly, all models are optimized on a part of the train set and best results are evaluated on the validation set. We use an Adam optimizer with an initial learning rate of 1e-4. Since this is a sequence task, models are trained and evaluated using the \emph{Concordance Correlation Coefficient} (CCC) metric.

\vspace{-0.2cm}
\section{Results}
\label{sec:results}

\begin{figure}[t]
	{
		\includegraphics[width=7cm]{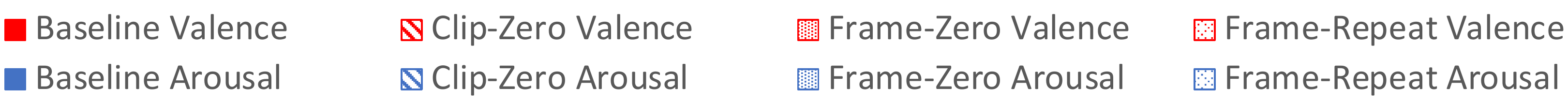}
		\label{fig:result_legend}
	}
	\subfigure[Clip-Zero]
	{
		\includegraphics[width=7cm]{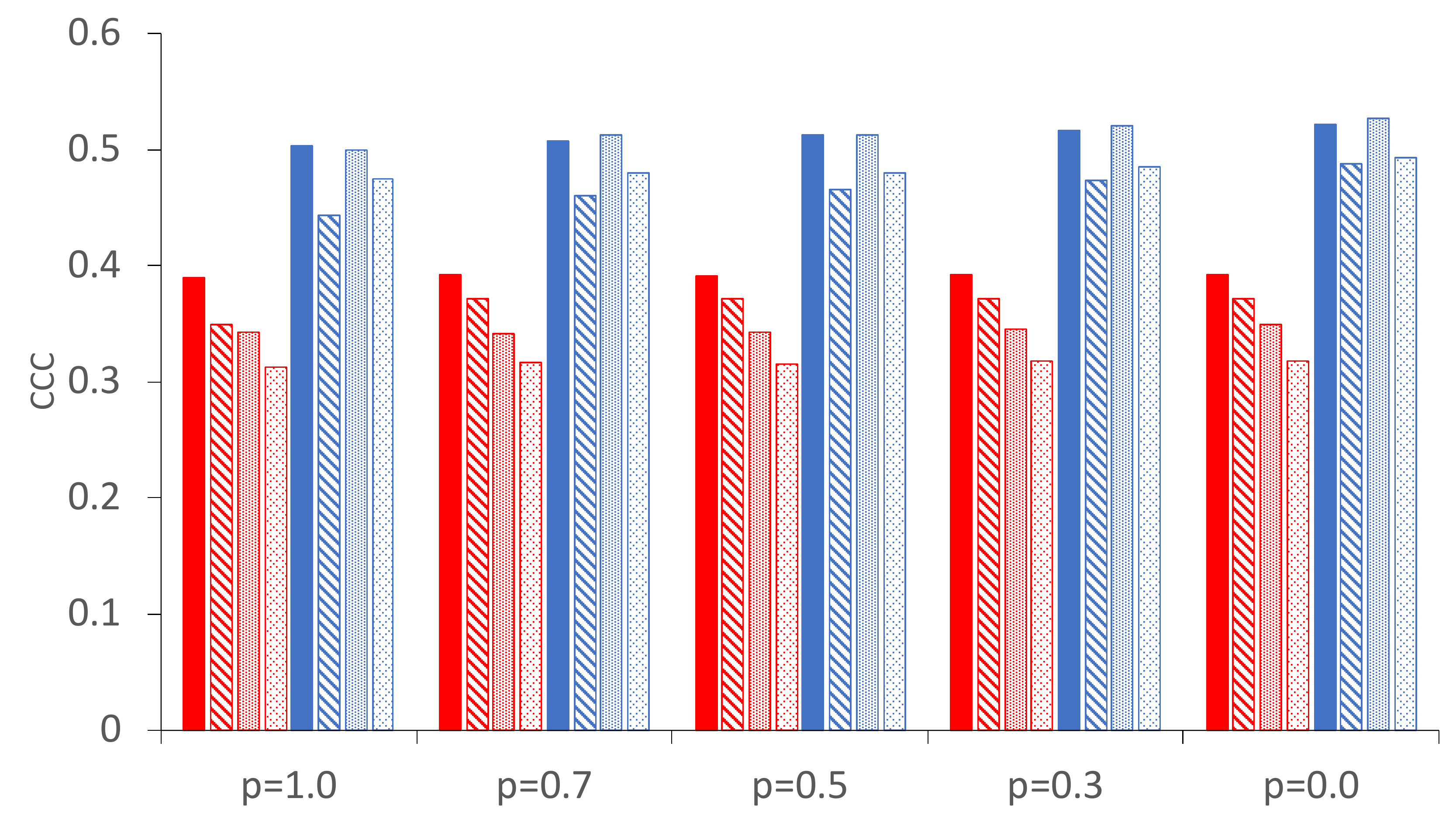}
		\label{fig:result_clip_audio}
	}
	\subfigure[Frame-Zero]
	{
		\includegraphics[width=7cm]{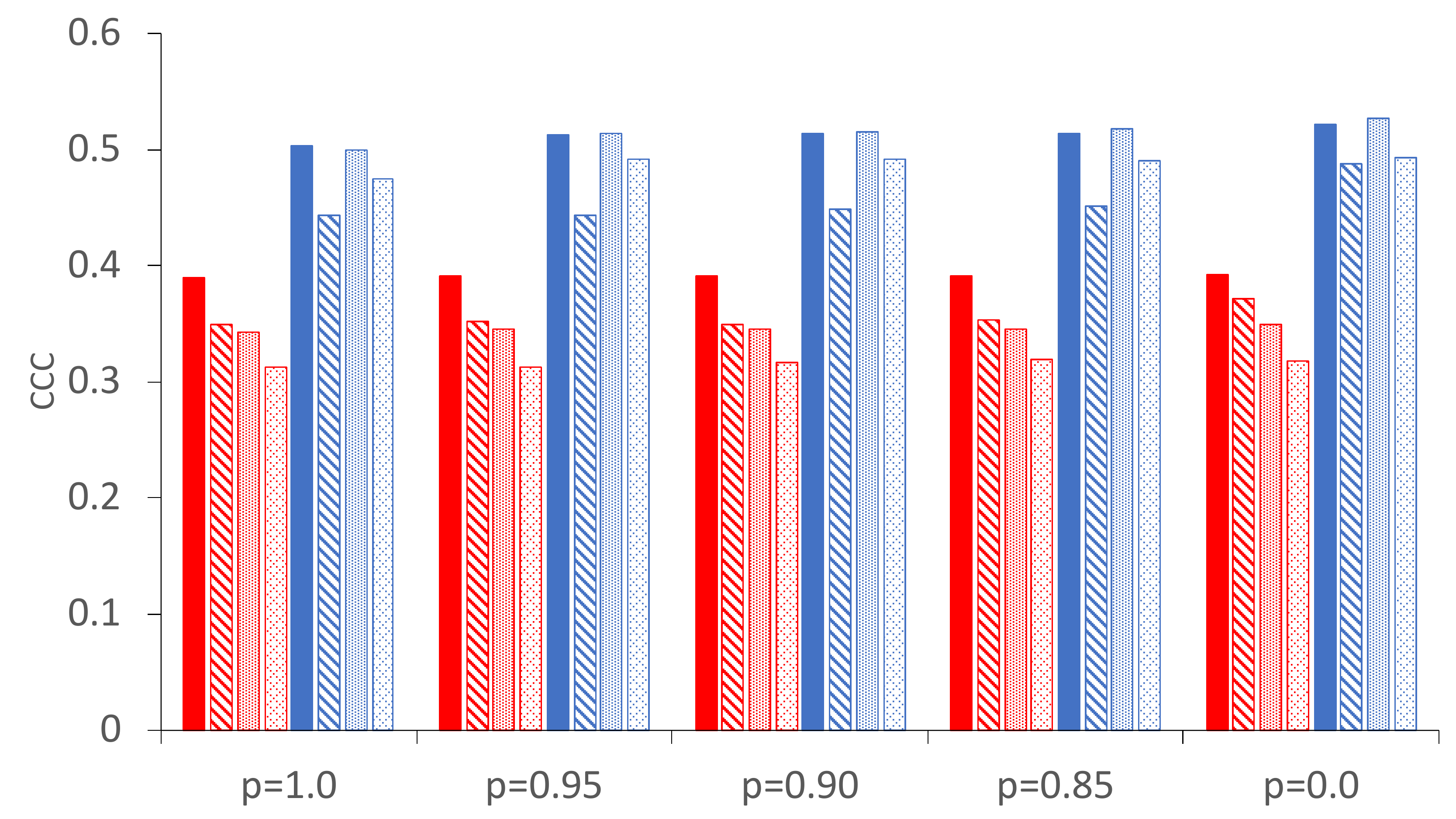}
		\label{fig:result_frame_zero_audio}
	}
	\subfigure[Frame-Repeat]
	{
	\includegraphics[width=7cm]{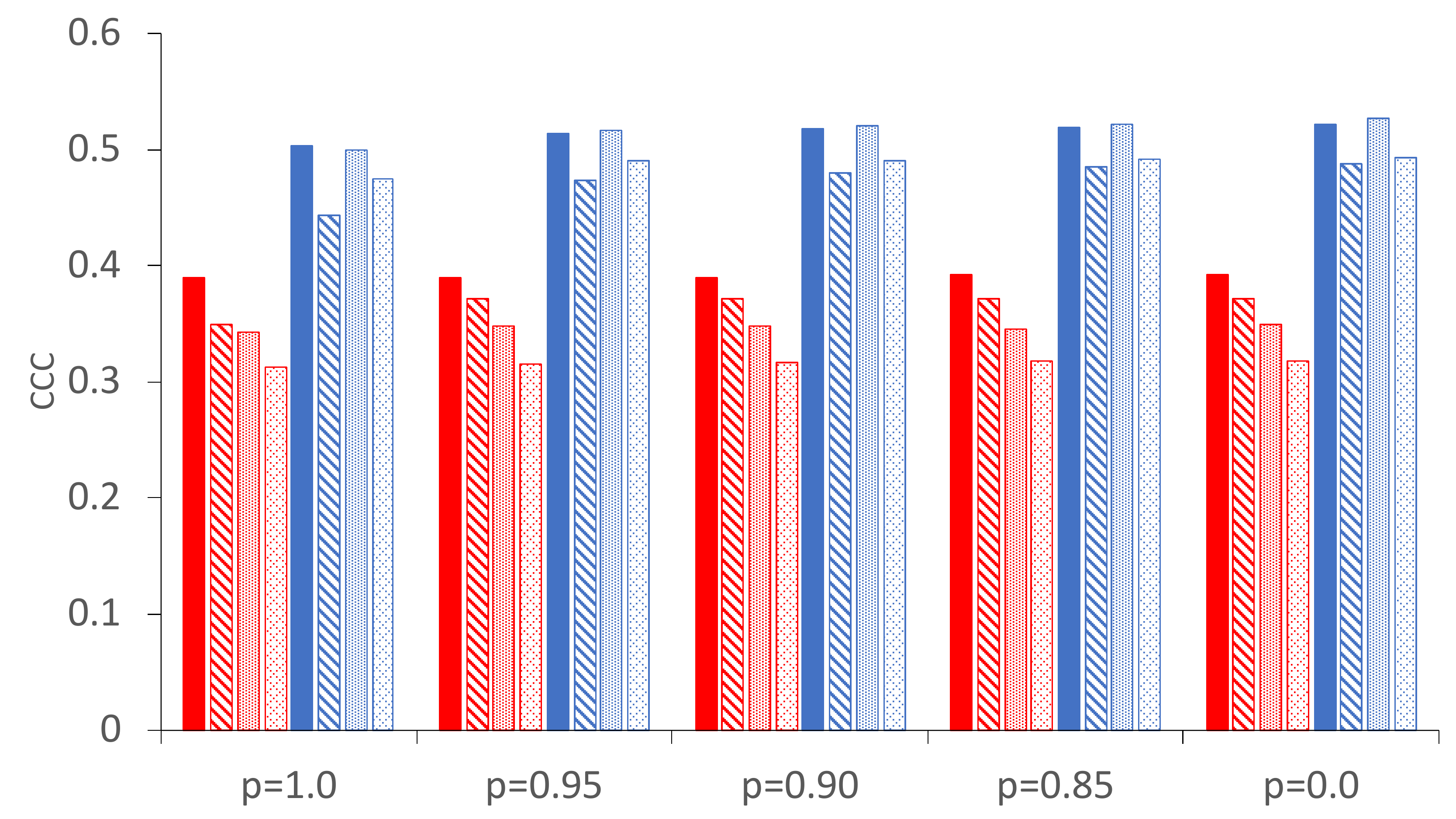}
	\label{fig:result_frame_repeat_audio}
	}
	\vspace{-0.2cm}	
	\caption{Results of training with missing data, evaluated with ablations of the audio modality at various probabilities for Clip-Zero, Frame-Zero and Frame-Repeat}
	\vspace{-0.4cm}
	\label{fig:results_audio}
\end{figure}

\begin{figure}[t]
	{
		\includegraphics[width=7cm]{figures/results_legend}
		\label{fig:result_legend2}
	}
	\subfigure[Clip-Zero]
	{
		\includegraphics[width=7cm]{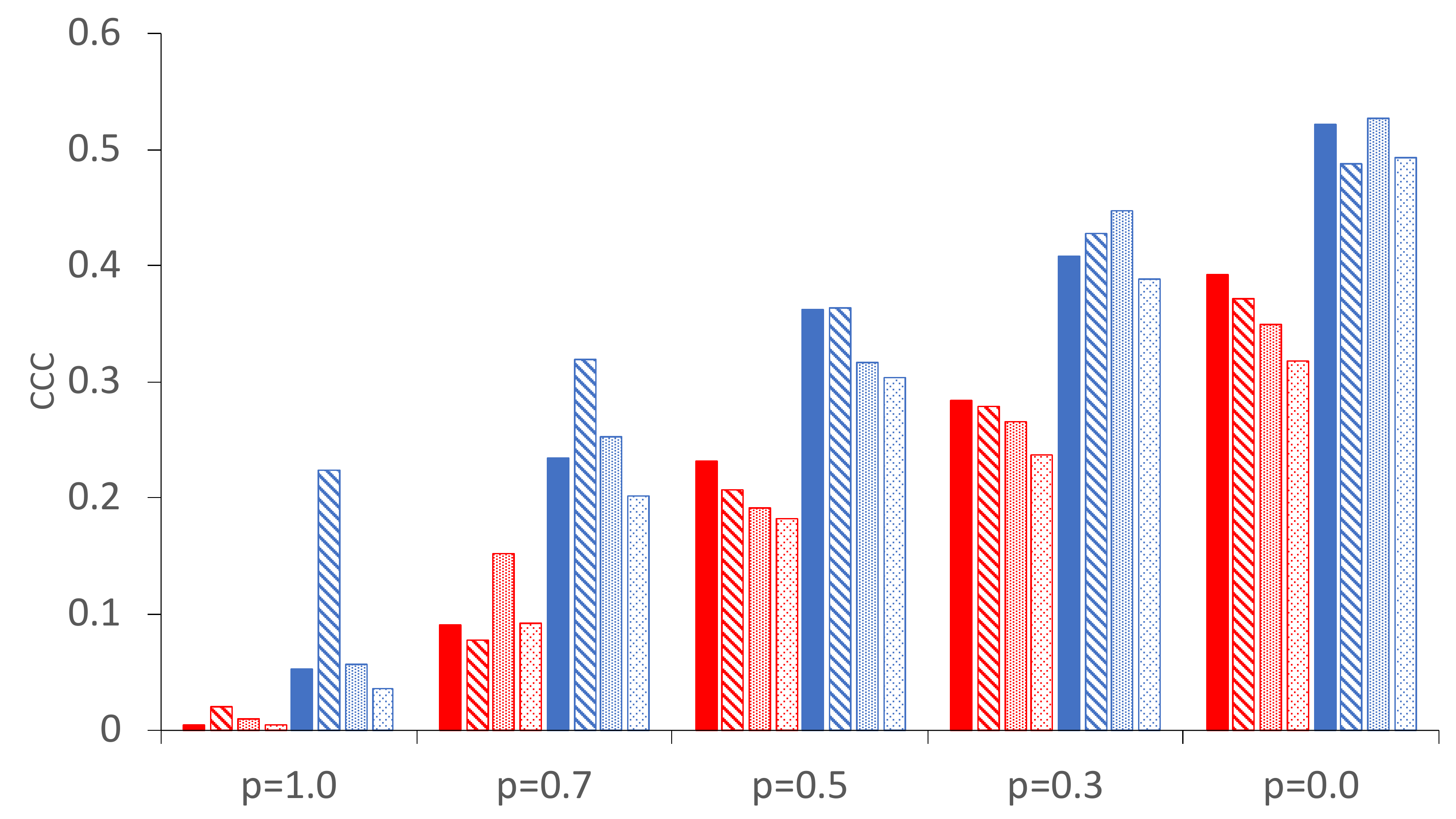}
		\label{fig:result_clip_video}
	}
	\subfigure[Frame-Zero]
	{
		\includegraphics[width=7cm]{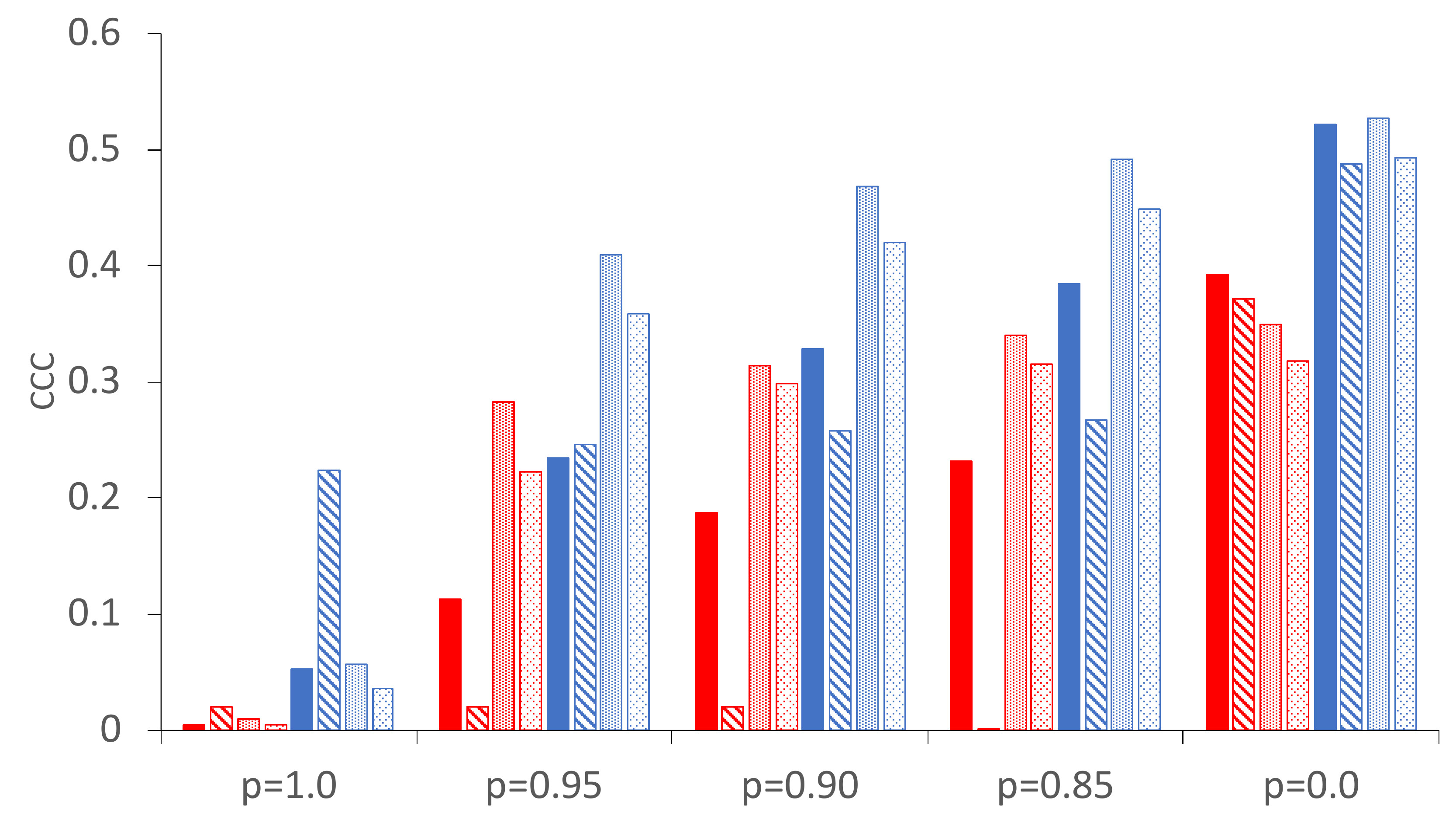}
		\label{fig:result_frame_zero_video}
	}
	\subfigure[Frame-Repeat]
	{
		\includegraphics[width=7cm]{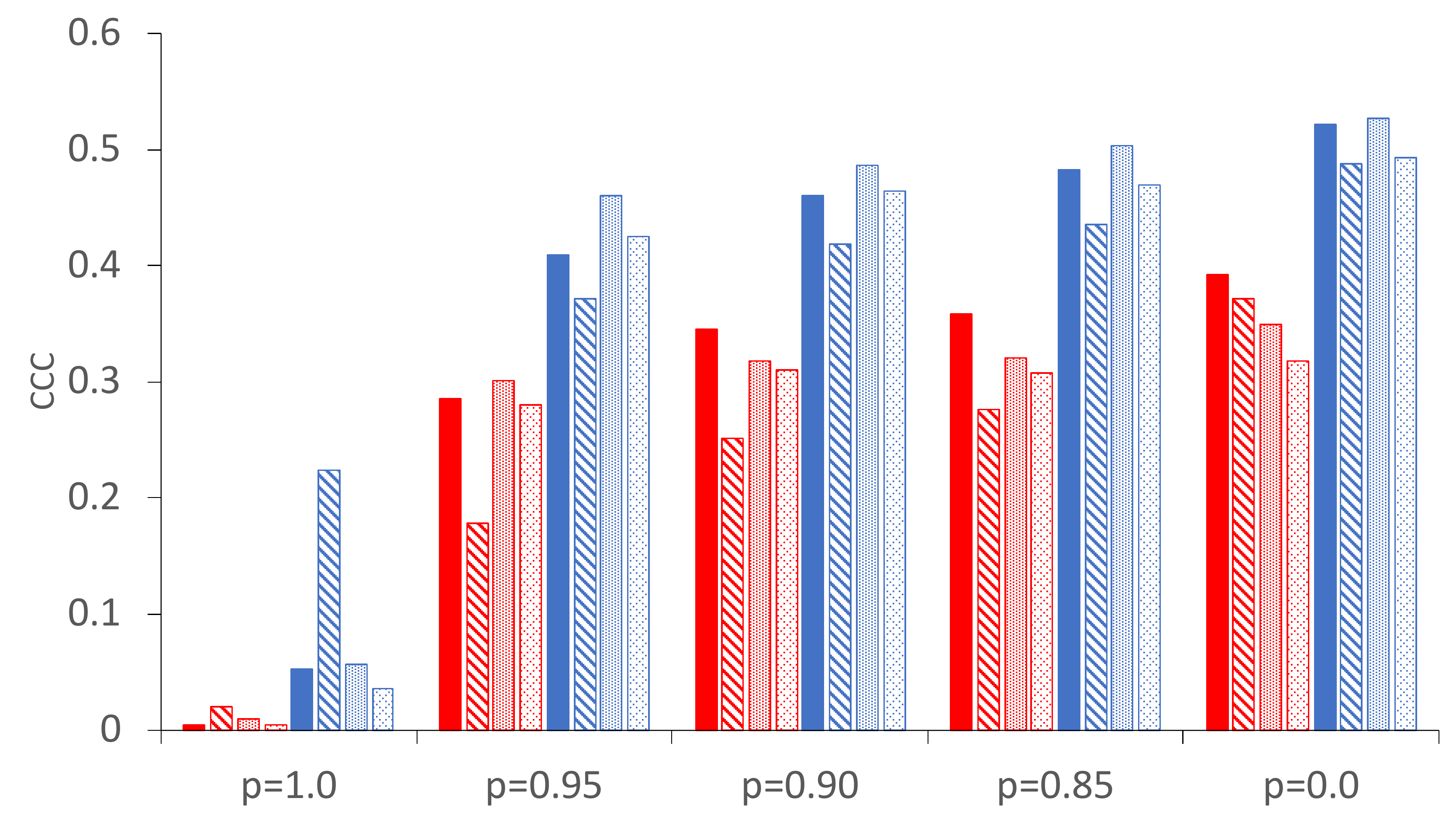}
		\label{fig:result_frame_repeat_video}
	}
	\vspace{-0.2cm}	
	\caption{Results of training with missing data, evaluated with ablations of the video modality at various probabilities for Clip-Zero, Frame-Zero and Frame-Repeat}
	\vspace{-0.4cm}
	\label{fig:results_video}
\end{figure}

To evaluate the proposed models, we propose various test conditions. Similar to the train conditions, we ablate the evaluation data inputs at  the clip level - \textbf{Clip-Zero} and frame level - \textbf{Frame-Zero} and \textbf{Frame-Repeat}, at various probabilities of missing data (Figure \ref{fig:results_audio} and Figure \ref{fig:results_video}). For Clip-Zero we evaluate using probabilities  $p \in$ {1.0, 0.7, 0.5, 0.3, 0.0} where the extremes correspond to the complete absence or complete presence of the modality. For Frame-Zero and Frame-Repeat we use more severe probabilities $p \in {1.0, 0.95, 0.90, 0.85, 0.0}$. 

We first observe that under ideal conditions (p = 0.0) the baseline model achieves CCC values of 0.393 and 0.522 for valence and arousal respectively. Under these conditions all proposed models are able to match baseline performance in most cases. Note that predicting valence suffers more than predicting arousal, indicating a greater dependence of valence on the visual modality.

\vspace{-0.25cm}
\subsection{Audio Ablations}
For audio ablations, all models in Figure \ref{fig:results_audio} show little variation in performance at various probabilities and various test conditions ($<$ 1\% absolute drop in performance). This indicates that the proposed models are more dependent on the rich video modality for the Aff-Wild2 database. While this may be particular to the Aff-Wild2, as shown by other studies, the ablation training methodology holds \cite{zhang2020m3t}. The proposed models cope with baseline performance even with half the video data removed. Note that a model trained only on the video stream, without cross-modal attention achieves CCC values of 0.381 for valence and 0.489 for arousal.

\vspace{-0.25cm}
\subsection{Video Ablations}
Perhaps the more interesting results are with evaluations on ablating the visual modality (Figure \ref{fig:results_video}). Evaluations with Clip-Zero (Figure \ref{fig:result_clip_video}) indicate that the model trained with Clip-Zero, overall handles ablations better than the baseline model. The performance is significantly better at high probabilities especially for arousal. Note that at $p=1.0$, the baseline model's performance is decreased by almost 100\%. In comparison  the Clip-Zero model performs similar to a model trained only on the audio-stream without cross-modal attention (CCC scores of -0.060, 0.284 for valence and arousal respectively) . This indicates that after training, the model is more dependent on audio cues, and automatically switches to inference using the available audio modality. Frame-Zero models show superior performance in certain scenarios for arousal when evaluated with clip-level ablations.

Evaluations with Frame-Zero (Figure \ref{fig:result_frame_zero_video}), Frame-Repeat (Figure \ref{fig:result_frame_repeat_video}) ablations indicate the necessity for training the model with matching conditions. Even at  $p=0.95$, where a large number of frames is missing the baseline model is able to recover significant performance (CCC-V=0.113, CCC-A=0.235 for Frame-Zero and CCC-V=0.285, CCC-A=0.410 for Frame-Repeat). But the proposed Frame-Zero and Frame-Repeat models do even better, and significantly outperform the baseline models for both attributes. Proposed models achieve gains of up to 17\% and quickly saturate to performance when evaluated at ideal conditions $p=0.0$. These results indicate the model is able to learn the expression from a few video frames in the sequence. The performance for Frame-Repeat evaluations is less pronounced than Frame-Zero evaluations but show similar trends. Overall these results show the generalization of the proposed training techniques across various test conditions.

\vspace{-0.25cm}
\section{Conclusion}
This study analyses the performance of a state-of-the-art transformer model, with attention, for audio-visual expression recognition in the wild. We show that the model suffers when input data is ablated. We propose a training strategy to handle missing input modalities. We randomly ablate visual cues by zeroing them out at the clip-level or zeroing or repeating them at the frame level. Other works have proposed ensemble approaches for expression recognition in the wild \cite{Gideon_2017, hu2017learning}. Our work can be viewed as a particular form of ensembling geared towards real world. Results indicate proposed models perform better than the baseline model when data is missing during evaluation without significant loss in performance under ideal conditions. These results indicate the superior generalization of the proposed models for various test cases addressing challenges for a real use case.

Our future work will explore further ablation strategies for training. In particular we will explore subclip level ablations where portions of the clip will be zeroed. Further different probability distributions can be utilized to ablate data at the frame level. We will also explore strategies where only the faces are ablated and the background scene information can be utilized rather than zeroing.

\bibliographystyle{ACM-Reference-Format}
\balance
\bibliography{reference}


\end{document}